\documentclass{article}

\usepackage{amsmath}
\usepackage{graphicx}

\newcommand\eqnn[2]{
        \begin{equation}#2\label{#1}
        \end{equation}}

\newcommand\elnn[2]{\begin{align}\begin{split}#2\label{#1}
        \end{split}\end{align}}

\begin{document}

\title{ General relativistic velocity: the alternative to dark matter 
}

\author{
        F. I. Cooperstock\thanks{{\tt cooperst@uvic.ca}}
        and S. Tieu\thanks{{\tt stieu@uvic.ca}}, \\
        Department of Physics and Astronomy\\
        University of Victoria \\
        P.O. Box 3055, Victoria, B.C. V8W 3P6 (Canada)
}

\maketitle

\begin{abstract}
We consider the gravitational collapse of a spherically 
symmetric ball of dust in the general relativistic weak gravity 
regime.
The velocity of the matter as viewed by external observers is compared to the velocity gauged by local observers. While the comparison in the case of very strong gravity is seen to follow the pattern familiar from studies of test particles falling towards a concentrated mass, the case of weak gravity is very different. The velocity of the dust that is witnessed by external observers is derived for the critically open case and is seen to differ markedly from the expectations based upon Newtonian gravity theory. Viewed as an idealized model for a cluster of galaxies, we find that with the general relativistic velocity expression, the higher-than-expected constituent velocities observed can be readily correlated with the solely baryonic measure of the mass, obviating the need to introduce extraneous dark matter.   
Hitherto unexplained and subject-to-reinterpretation 
astrophysical phenomena could also be considered within this context. 
It is suggested that an attempt be made to formulate an experimental 
design at smaller scales simulating or realizing a collapse with the 
aim of implementing a new test of general relativity.

\end{abstract}



Recently \cite{CT4}, \cite{CT5}, we demonstrated that general 
relativity could account for the flat galactic rotation curves, the observation of the essentially constant velocities of the stars in the galaxies out to their extremities,  
without the requirement for the conventionally demanded vast 
reservoirs of exotic dark matter. \footnote{Various critics have claimed that 
our results stemmed from a singular surface layer of mass but we have 
shown that the singularity actually represents the benign incorporation of a 
discontinuity in density gradient. As well, a different approach 
\cite{BG} largely supported our central thesis.
}
 We have seen that general 
relativistic nonlinearities can play an important role even in the context of weak gravity.

To fortify our claim, and to attack frontally the issue of higher-than-expected 
velocities on the basis of Newtonian gravity and the observed
 matter in clusters of galaxies, we now consider a 
different problem, the gravitational collapse \footnote{
	The expression ``gravitational collapse'' has come
	to mean the total crunching of matter under gravity
	but we are using it here in the general sense of
	material flowing inward under gravity.
}
of a spherically symmetric sphere of dust in the phase where a ball 
structure is evident with external vacuum. While the interest in the 
collapse problem in the past has focused upon strong gravity leading 
to singularity formation (see, e.g. \cite{coop})\footnote{
	The essential point in that paper was the implementation of 
	pressure during the late stages of gravitational collapse. 
	The maintenance of the dust equation of state is totally 
	unrealistic physically when the concentration of material 
	mounts dramatically in the approach to zero volume.
},
we consider here the \textit{weak} gravity regime, long before any 
singularity could be formed \footnote{
	Part of the motivation for analyzing the present problem stems 
	from the fact that singularities cannot be raised here as 
	an issue. 
} 
This could be viewed as a special case model, albeit highly idealized, of a 
cluster of galaxies in evolution where in the normally unsymmetric 
case, the component velocities have been observed to have unusually 
high velocities \footnote{
When we speak of ``high'' velocities in this context, we mean high compared to the expectation of Newtonian gravity but still very much smaller than the speed of light. Also, when we speak of ``general relativistic velocity'', velocity governed by Einstein's theory of gravity, such velocity can take on the whole range of values, but in this paper, the emphasis is on the cases where this velocity is much less than the velocity of light. This expression should not be confused with the familiar expression ``relativistic velocity'' of special relativity where the reference is to velocity approaching the speed of light.
}
according to Newtonian gravity. Reportedly, this 
phenomenon was the historic origin of the dark matter hypothesis, 
advanced in the 1930's by Zwicky, in an effort to explain the 
high velocity observations within the context of Newtonian gravity theory. In this 
paper, we show that in this idealized model, such velocities can be 
accounted for in principle using general relativity in the absence of extraneous 
assists from dark matter. 

We first consider the treatment in \cite{LL} (henceforth referred to 
as ``LL'') of the familiar Schwarzschild solution, the spherically 
symmetric vacuum gravitational field for a spherically symmetric mass $m$. A spherically symmetric metric can be expressed in generality in spherical polar coordinates in the form
 \footnote{
	We choose units where $c=G=1$.
}
\eqnn{Eq1d}{
	ds^2= e^{\nu(r,t)} dt^2 - e^{\Lambda(r,t)} dr^2
		- r^2(d\theta^2 +\sin^2\theta d\varphi^2)
.}
In the case of vacuum, the metric functions are readily found by solving the Einstein field equations and reflecting the intrinsically static nature of the spherically symmetric vacuum solution, can be expressed in the time-independent form
\eqnn{Eq2d}{
	e^{\nu}= 1-\frac{2m}{r},\quad\quad
	e^{\Lambda}= \left(1-\frac{2m}{r}\right)^{-1}
.}
This is generally referred to as the Schwarzschild metric. LL effect the transformation to comoving
\footnote{
In these coordinates, an observer being at $R=constant$ is in free-fall.
}
synchronous \cite{LL} coordinates $R, \tau$ as
\footnote{
$\theta$ and $\phi$ are left unaltered.
}
\elnn{Eq3d}{
	\tau&=t+\int \frac{f(r)}{1-\frac{2m}{r}} \,dr \\
	R&=t+\int
		\frac{1}{ f(r)\left(1-\frac{2m}{r}\right)} \,dr
}
with $f(r)$ chosen as
\eqnn{Eq4d}{
	f(r)=\sqrt{ \frac{2m}{r}}
}
yielding the simple relationship between the coordinates
\eqnn{Eq5d}{
	r=\left( \frac{3}{2} ( R-\tau) \right)^{2/3} ( 2m)^{1/3}.
}
The singularity issues encountered in (\ref{Eq1d}),(\ref{Eq2d}) at 
$r=2m$ are thus alleviated. The metric 
in these new comoving $(R,\tau)$ coordinates is then expressed as 
\eqnn{Eq6d}{
	ds^2= d\tau^2
	-\frac{dR^2}{\left(\frac{3}{2(2m)}(R-\tau)\right)^{2/3}}
	-\left(\frac{3}{2}(R-\tau)\right)^{4/3}(2m)^{2/3}
		(d\theta^2 +\sin^2\theta d\varphi^2)
}
which is now seen to be time-dependent. 
The strong gravity regime evolves as the proper time $\tau$ 
approaches $R$ and the singularity emerges at $R=\tau$.  

Our focus will be on the weak gravity regime where $R>>\tau$ for all $R$. This 
translates to $r>>2m$ for all $r$ in the $(r,t)$ frame.
	The time coordinate $t$ measures time read by the 
	asymptotic observer. The standard general relativistic treatment concentrates upon the regime of strong gravity where the difference in perception of the proper velocity of a freely falling test particle as measured by the local observer in comparison to the measurement of the velocity by the asymptotic observer becomes particularly significant(see \cite{LL}).

In generality the proper radial velocity of a freely falling test particle cannot be 
evaluated in the $(R,\tau)$ 
coordinates. This is  because $R$ is constant for any given particle in this 
frame and hence the radial velocity is always zero in this comoving 
frame. For the required ingredients, LL use the solution of the 
radial geodesic equation for a freely falling test particle in the 
usual Schwarzschild coordinates $(r,t)$ as employed in  (\ref{Eq1d}),(\ref{Eq2d}) which are suitable for this 
purpose. The geodesic solution for $dr/dt$ and the metric 
coefficients $g_{00}$ and $g_{11}$ of  (\ref{Eq1d}) \footnote{
	$(x^0,x^1,x^2,x^3)= (t,r,\theta, \phi)$
}
are used to evaluate the proper radial velocity
\eqnn{Eq8d}{
	v= -\sqrt{\frac{-g_{11}}{g_{00}}}\frac{dr}{dt}
}
This equals $\sqrt{2m/r}$ in magnitude 
for particles released from rest at infinity and is seen to approach 1, the speed of light, as $r$ approaches $2m$. (The rest release point 
$r_0$ in LL Eq.(102.7) is taken to be infinite here.) 
However, for asymptotic observers who reckon radial distance and time increments as $dr$ and $dt$, the measured velocity is
\eqnn{Eq8e}{
\frac{dr}{dt} = -\left(1-\frac{2m}{r}\right)\sqrt{\frac{2m}{r}} 
}
which approaches zero in the very strong gravity regime as $r$ approaches $2m$, in stark contrast to the proper radial velocity. However, for weak gravity which is our focus in this paper, the $(1-2m/r)$ factor in (\ref{Eq8e}) is approximately 1 and the local proper and asymptotic measures of velocity are approximately equal in the value $-\sqrt{2m/r}$. This justifies the neglect of general relativity in the context of weak gravity for the case of motion in spherically symmetric vacuum.
  
Interesting new developments ensue when we turn to dust collapse.
As with the vacuum case, LL choose comoving coordinates for dust 
collapse and structurally as in (\ref{Eq6d}), express the metric as 
\eqnn{Eq9d}{
	ds^2= d\tau^2 - e^{\lambda(\tau,R)} dR^2
		- r^2(\tau,R)(d\theta^2 +\sin^2\theta d\varphi^2)
}
The four non-vanishing Einstein field equations are extremely
non-linear:\footnote{
	A dot denotes the partial derivative with respect
	to $\tau$ and a prime denotes the partial derivative
	with respect to $R$.
} 
\eqnn{Eq10d}{
	-e^{-\lambda} (r')^2 + 2r\ddot{r} +{\dot{r}}^2 +1 =0,
} 
\eqnn{Eq11d}{
	-\frac{e^{-\lambda}}{r} \left(
		2r'' - r' \lambda' \right)
	+\frac{\dot{r}\dot{\lambda}}{r}
	+\ddot{\lambda}
	+\frac{\dot{\lambda}^2}{2}
	+\frac{2\ddot{r}}{r}
	=0
}
\eqnn{Eq12d}{
	-\frac{e^{-\lambda}}{r^2} \left(
		2rr'' + (r')^2 - r r' \lambda' \right)
	+\frac{1}{r^2}\left( r\dot{r}\dot{\lambda} +\dot{r}^2+1\right)
	=8\pi \rho
}
\eqnn{Eq13d}{
	2(\dot{r})' - \dot{\lambda} r' = 0
}

It is remarkable that for such a complicated non-linear set of equations, at 
least a part of the solution should be as simple as it is:
\footnote{
$E(R)$ and $F(R)$ are functions of integration.  
}
\eqnn{Eq14d}{
	e^\lambda = \frac{(r')^2}{1+E(R)}
}
\eqnn{Eq15d}{
	\dot{r}^2=E(R) + \frac{F(R)}{r}
}
\eqnn{Eq16d}{
	r = \left(\frac{9F}{4}\right)^{1/3}
		( R - \tau )^{2/3}
}
where in (\ref{Eq16d}), we have chosen the part of the three-stage solution \cite{LL} for the case 
where $E(R)$ in (\ref{Eq15d}) is taken to be zero, the particles being released from 
rest at infinity in the infinitely distant past. \footnote{
	There are two other cases for non-zero $E$, the bound and the unbound cases, which are familiar in concept from classical mechanics. These
	solutions are expressed in parametrized form in \cite{LL}. Here, we focus upon the simplest critically open case. 
} 
\footnote{Note that in general, one can choose an arbitrary function 
${\tau}_0(R)$ in place of $R$ in the $(R-\tau)^{2/3}$ factor in 
(\ref{Eq16d}). We have chosen ${\tau}_0(R)=R$ to mesh smoothly with the 
chosen vacuum Schwarzschild solution in comoving coordinates (\ref{Eq5d})which holds in the vacuum 
region surrounding the ball of collapsing dust.
}
For all three cases, positive, negative or zero $E$, the density $\rho$ is 
incorporated into the solution as 
\eqnn{Eq17d}{
	8\pi  \rho = \frac{F'}{r' r^2}.
}

From (\ref{Eq17d}), a simple  integration (see \cite{LL}) shows that the mass $M(R)$ within the radial coordinate $R$ is
\eqnn{Eq18d}{
	M(R)= F(R)/2
}
and thus the entire mass $M$ is given by $M(R_0)$ where $R_0$ is the 
outer comoving radial coordinate of the dust ball. \footnote{
	Note that there is no problem in matching the interior dust 
	solution with that of the exterior vacuum Schwarzschild 
	solution and no reasonable argument can be raised as to the 
	presence of a surface layer. A well-known example of 
	interior-to-exterior matching with spherical symmetry is 
	that of Schwarzschild's static constant density ball with 
	pressure matched to the exterior vacuum solution.
} 


Our focus here is upon the radial dust velocity measured by distant ``rest'' (i.e. \textit{non}-comoving) observers. As with the vacuum case, we must choose new coordinates
for the 
evaluation of the radial velocity because the $R$ coordinate is 
constant for any given dust particle.  As discussed above, for the dust case, we continue to use the approach taken by LL for vacuum and
evaluate this radial velocity $dr/dt$ in the Schwarzschild-like $(r,t)$ coordinate frame. However, unlike the case of vacuum, it is unnecessary to solve the geodesic equations at this point because the motion of the dust medium has already been solved in the comoving 
frame. It is this motion that is of concern to us.\footnote{
	Actually, the motion is also geodesic here since there is
	no pressure.
}
What is required is to re-express the solution in Schwarzschild-like $(r,t)$
coordinates. 
\footnote{Within the dust ball, the coordinates used cannot be expressly 
Schwarzschild coordinates as used with explicit time-independence in  
(\ref{Eq2d})because the field within the dust is intrinsically 
dynamic. However, the essential metric structure is the same. As well, by making $r^2$ the coefficient of the angular part of the metric, the circumference of a ring of particles at $r$ assumes the familiar flat-space value $2{\pi}r$ for both the proper measure and for the measure as judged by distant observers.
}
For consistency with the solution form of (\ref{Eq16d}) and to maintain maximum available generality, we choose the general form of transformation with arbitrary functions $p(r,t)$ and $q(r,t)$,
\elnn{Eq20d}{
	\sqrt{F}R= p(r,t),\sqrt{F}\tau&= q(r,t)
}
with the constraint
\elnn{Eq20e}{
p(r,t)-q(r,t)=(2/3)r^{3/2}
.}
From (\ref{Eq20e}), we see that \footnote{
Here and in any subsequent appearances of $p(r,t)$ and $q(r,t)$, a dot on these functions denotes the partial derivative with respect
	to $t$ and a prime on these functions denotes the partial derivative
	with respect to $r$.
}
\elnn{Eq20f}{
p'(r,t)-q'(r,t)=r^{1/2},\dot{p}(r,t)=\dot{q}(r,t). 
}

We take differentials of (\ref{Eq20d}), and with (\ref{Eq20e}),(\ref{Eq20f}),  solve for $dR$ and $d\tau$. These differentials are substituted into (\ref{Eq9d}) to derive the normal form of the metric in Schwarzschild-like coordinates $(r,t)$ with terms of the form $g_{00}dt^2$ and $g_{rr}dr^2$, as well as an undesired cross-term of the form $2g_{0r}drdt$. This cross-term must vanish to mesh with the exterior Schwarzschild metric at the vacuum interface and maintain the useful Schwarzschild-like non-rotating form within the ball.  This metric form includes a yet-to-be-determined $p'(r,t)$ which we set by making $g_{0r}=0$ yielding
\elnn{Eq20fa}{
p'=\frac{ (\frac{3R\sqrt{F}\alpha}{2r} +\sqrt{r}\beta)}{(\alpha + \beta)(1-\beta^2)}
}
where
\elnn{Eq20fb}{
\alpha=\frac{rF'}{3F} = \frac{r M'(R)}{3M(R)} \\
	\beta= \sqrt{\frac{F}{r}} = \sqrt{ \frac{2M(R)}{r} }
.}
Also required in the calculation for $p'$ is $e^\lambda$ which, from (\ref{Eq14d}), is equal to $(r')^2$ for $E=0$. In turn, this requires $r'$ which is computed from (\ref{Eq16d}) yielding
\elnn{Eq20fc}{
r'= \alpha + \beta.
}

Since the $R$ coordinate is comoving with the matter, we express the condition for the radial motion of the particles by taking differentials of the first of (\ref{Eq20d}) and setting $dR=0$:
\footnote{
For motion in this spherically symmetric study, $d\theta$=$d\phi$=$0$ as well.
}
\elnn{Eq20g}{
p'(r,t)dr + \dot{p}(r,t)dt=0 
}
from which we find the form of the radial velocity of the particles as witnessed by external observers
\elnn{Eq20h}{
dr/dt =-\dot{p}(r,t)/p'(r,t). 
}
To solve for $\dot{p}(r,t)$, we first apply $\partial/\partial{t}$ to (\ref{Eq17d}):
\elnn{Eq20i}{
8\pi\frac{\partial\rho}{\partial{t}}= \frac{ F'^2\left(\frac{\alpha}{F}+ \beta(\frac{ F''}{ F'^2}-\frac{1}{2F})\right)         \dot{p}}{r^2(\alpha + \beta)^2 (\frac{3R\sqrt{F}\alpha}{2r} +\sqrt{r}\beta)}
}
The derivation of (\ref{Eq20i}) made use of (\ref{Eq20fc}) 
\elnn{Eq20k}{
\frac{\partial{(\alpha +\beta)}}{\partial{t}}= \left[\frac{F'}{2\sqrt{Fr}} +\frac{r}{3}(\frac{F''}{F}-\frac{F'^2}{F^2})\right]\frac{\partial{R}}{\partial{t}}
}
(where(\ref{Eq20fb}) has been used) 
and the elimination of $\frac{\partial{R}}{\partial{t}}$
using 
\elnn{Eq20l}{
\dot{p} =\left[\frac{F'R}{2\sqrt{F}} +\sqrt{F}\right] \frac{\partial{R}}{\partial{t}}
}  
which follows from the partial differentiation with respect to $t$ of the first of (\ref{Eq20d}). 
Finally, using (\ref{Eq20fa})and(\ref{Eq20i}) in conjunction with (\ref{Eq20h}) (and with a cancellation of the factor $(\frac{3R\sqrt{F}\alpha}{2r} +\sqrt{r}\beta)$), we find
\footnote{
Note that a lengthy calculation with the metric components in the $(r,t)$ frame in conjunction with  (\ref{Eq8d}) yields the proper radial velocity, the velocity measured by local observers, with value $\sqrt{\frac{F}{r}}$. This coincides with $\dot{r}= \frac{\partial{r}}{\partial{\tau}}$. However for observations by external observers, the measured velocity is $dr/dt$ as given by (\ref{Eq20m}) \cite{LL}.
} 

\elnn{Eq20m}{
        \frac{dr}{dt}=-\frac{(\alpha+\beta)(1-\beta^2)}{8\pi r^2 \rho^2}\left[\frac{\alpha}{F} + \beta\left(\frac{F''}{(F')^2} - \frac{1}{2F}\right)\right]^{-1}\frac{\partial\rho}{\partial t}
}

This is the key equation. The complexity of this velocity expression as computed by observers external to the distribution of matter is in very sharp contrast to the simplicity of the proper velocity form $\beta=\sqrt{\frac{F}{r}}$ as witnessed by local observers. However, \textit{it is the former that is relevant for astronomical observers.} For local observers, it is the mass, $ F(R)/2 = M(R)$, at radii within the point of interest that determines the velocity, the situation as in Newtonian gravity. However, we see in (\ref{Eq20m}) that within the context of general relativity, the external observers ascribe velocity based on additional factors, the reciprocal of the local density squared and its time rate of change (also expressible as the time rate of change of reciprocal density), the gradient of the mass within the radius in question, $M'(R)$ as well as well as its gradient, $M''(R)$.

We also see that in the limit of very strong gravity, with $\beta$ approaching $1$, the situation is the same as we witnessed in vacuum: the local observers see the velocity approach $1$ whereas the external observers see the velocity approach $0$. 

However for weak gravity with $\beta <<1$, the vacuum and dust comparison is very different. While the local and asymptotic velocity measures for observers plotting freely falling test particles in vacuum in the field of a concentrated mass are approximately the same, namely $\beta$, the corresponding velocities for local and asymptotic measure in the case of dust are very different in general: the velocity is simply $\beta$ for the local measure whereas the asymptotic measure is given by the rich expression (\ref{Eq20m}) with $1-\beta^2$ approximated by $1$. Indeed, given the complexity of the form of $dr/dt$ in (\ref{Eq20m}), it would be a very special occurrence for $dr/dt$ to have the value $\beta$. Thus, when astronomers witnessed with consternation velocities greater than $\beta$ in galactic clusters, they should have been more reasonably surprised had they witnessed $\beta$ velocities.

As an application of (\ref{Eq20m}), we first consider the astrophysical realm. Most of the gravity in the universe is weak gravity where Newtonian theory has been deemed to be perfectly adequate. Thus, when galaxies in clusters, with gravity found to be weak, were seen to have velocities exceeding $\beta$, dark matter was introduced as the necessary mass booster to align the observations with enlarged $\beta$. Newtonian theory formed the basis for the calculations. However, we have seen that general relativity, which is almost universally accepted as the preferred theory of gravity, actually predicts velocities that have elements beyond $\beta$ even when the gravity is weak. A key point is that the nonlinearities of general relativity play an important role in this problem, leading to expressly non-Newtonian behaviour, even though the gravity is weak.

 For the Coma Cluster of galaxies, the ratio\footnote{
The key equation (\ref{Eq20m}) is most conveniently expressed in terms of $r$ and $R$. While $r$ has a direct measurable connection to the source in that $4\pi r_0^2$ is the surface area of the ball, there is a great deal of arbitrariness attached to the choice of the numerical value of $R_0$. From the transformation equations, we see that different settings of zero value for the clocks will change the number attached to $R_0$. For the purpose of normalization to enable numerical calculations, we must choose $R_0=r_0$ so that the average density of the ball calculated using $r_0$ will equal that using $R_0$. The zero setting of the clock is adjusted to assure this equality. With the average densities normalized, the explicit calculation of $dr/dt$ can proceed.
}    
$2M(R_0)/r_0$ is of order $10^{-4}$ assuming the existence of dark 
matter and of the order $10^{-5}$ by not assuming any dark matter. 
The gravity is indeed very weak in this source for the kind of applications under 
consideration. 
Thus as a test model, we consider an idealized Coma Cluster of galaxies, one of spherical symmetry with the velocities as reported in \cite{Hughes}. At a radius of 1 Mpc, the total cluster mass, including dark matter, is given as $6.2\times 10^{14} M_{\odot}$,  with the 13\%-17\% portion being normal baryonic matter. Within a radius of 3 Mpc, the total mass is reported to be $1.3\times 10^{15} M_{\odot}$, 
with the normal luminous matter portion within the wide range of 20\%-40\%.

\begin{figure}
\begin{center}
\includegraphics[width=4in]{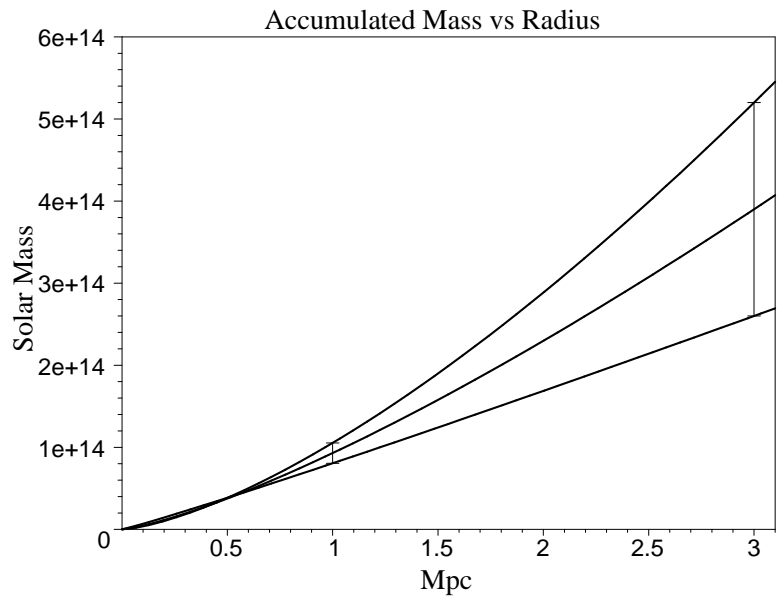}
\end{center}
\caption{The upper, middle and lower limits of mass accumulation
        are described by the functions,
        $F= 6.641 \times 10^{-16} R^{1.453}$,
        $F= 1.244 \times 10^{-12} R^{1.305}$ and
        $F= 2.531 \times 10^{-7}  R^{1.066}$ respectively.
}
\label{fig:accmass}
\end{figure}

\begin{figure}
\begin{center}
\includegraphics[width=4in]{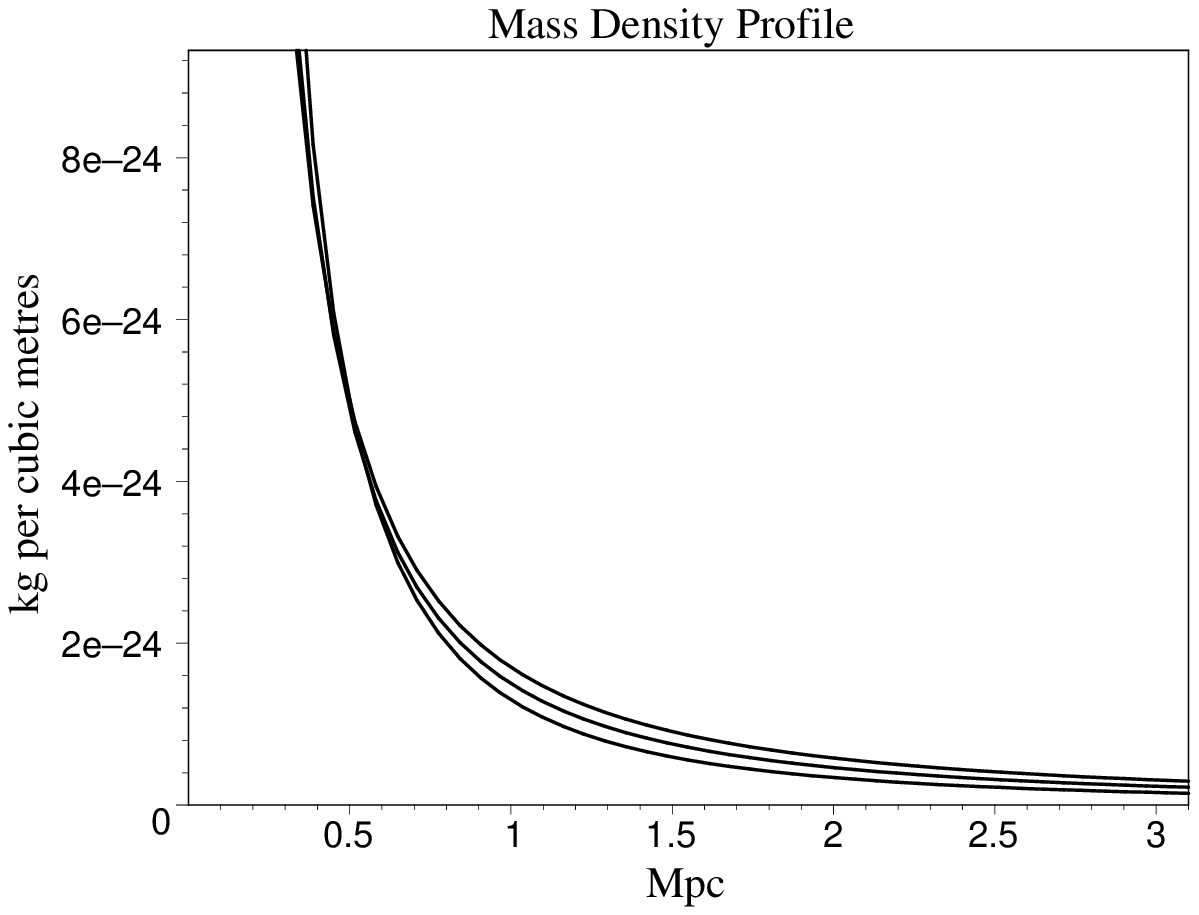}
\end{center}
\caption{From the three functions,
        $F= 6.641 \times 10^{-16} R^{1.453}$,
        $F= 1.244 \times 10^{-12} R^{1.305}$ and
        $F= 2.531 \times 10^{-7}  R^{1.066}$, we can derive
        the mass density as shown in the graph.
}
\label{fig:density}
\end{figure}

It is simple to fit these data with an accumulated mass function
\elnn{Eq30d}{
        F(R) = k_1 R^{k_2},
}
($k_1$, $k_2$ constants) 
as shown in Figure \ref{fig:accmass}.
With $F(0)=0$ from (\ref{Eq30d}), we are assured that there is no singularity at the origin \cite{LL}.
Using (\ref{Eq30d}) in (\ref{Eq17d}), we derive the density profile for the distribution.
The graph of the densities for the two extremes of the uncertainty range
and the average is shown in Figure \ref{fig:density}.  

The velocity associated with each $F(R)$ is given by (\ref{Eq20m}) where we can set the ``boosted'' velocity as
\elnn{Eq31d}{
dr/dt =-n\beta
}
where $\beta$, as throughout the paper, is composed from the baryonic mass alone and $n$ is the ``booster'' number to bring $dr/dt$ to the observed level of velocity.    
Assuming the baryonic mass is 20\%, 30\% and 40\% of $1.3\times 10^{15} M_{\odot}$, we
find that the boost factors $n$ are 2.23, 1.82 and 1.58, respectively.
Applying this to (\ref{Eq20m}),
we can solve for
$\partial\rho/\partial t$, the sole unknown factor.  The results are:  $2.13\times 10^{-41} $kg/m$^3$/sec,
$2.62\times 10^{-41} $kg/m$^3$/sec and $3.02\times 10^{-41} $kg/m$^3$/sec,
respectively.
Rates such as  $10^{-41}$kg/m$^3$/sec are quite reasonable as over a period of one billion years,
the density would grow by $10^{-25}$ kg/m$^3$, hence roughly doubling the value of the present density.

In this example, we see adequate scope to explain the observed velocities within the framework of general relativity without the requirement of any extraneous dark matter. The new elements of local density, its time rate of change, the gradient of the mass interior to the observation point as well as its gradient are additional factors that ultimately determine the net observed velocity of the matter by external observers.
While this is an idealized case of perfect spherical symmetry, it 
would seem reasonable to expect comparable effects for non-spherical 
accumulations of freely-gravitating collections of bodies as we have 
in clusters of galaxies.  Had Zwicky made this calculation 70 years 
ago, he might have come to very different conclusions.

Clearly there is considerable further analysis ahead. This paper has only dealt with the simplest case $E(R)=0$. The positive and negative cases for $E$ offer greater freedom of expression.  Ultimately, the ideal would be to formulate the equivalent effects of general relativity as applied to \textit{chaotic} weakly- gravitating systems. For this, the general relativistic equivalent of the virial theorem is called for. As well, there is the issue of the interpretation of lensing as a mechanism for the deduction of mass. The subtleties of general relativistic weak gravity that we have found in the present work must now be directed to the consideration of lensing.

It must be stressed that as before \cite{CT4}, \cite{CT5}, we are 
witnessing here the power of the nonlinearities inherent in general 
relativity in the context of weak gravity to effect very significant 
changes relative to the results expected on the basis of Newtonian 
theory. It suggests that hitherto unexplained astrophysical phenomena 
be re-considered on the basis of the application of general 
relativity to weak-field gravity.  Indeed the present case also 
suggests that an attempt be made to formulate an experimental design 
at smaller scales simulating or realizing a collapse with the aim of 
implementing a new test of general relativity.


\vskip 0.125in
{\bf Acknowledgments}
\vskip 0.125in 
 This work was supported in part by a grant from 
the Natural Sciences and Engineering Research Council of Canada.

\end{document}